\documentclass[a4paper, pra, twocolumn,superscriptaddress,showpacs]{revtex4}

\usepackage{amssymb}
\usepackage{amsmath}
\usepackage{epsfig}
\usepackage{color}
\usepackage{graphics, graphicx}
\usepackage{bbold}
\usepackage{psfrag}
\usepackage{mathcomp}
\usepackage{subfigure}
\usepackage{verbatim}
\usepackage{color}
\usepackage[colorlinks,citecolor=blue]{hyperref}
\def\cp#1{\mathbf{#1}}

\begin{document}

\date{\today}
\title{Universal trimers emerging from a spin-orbit coupled Fermi sea}
\author{Xingze Qiu}
\affiliation{Key Laboratory of Quantum Information, University of Science and Technology of China, CAS, Hefei, Anhui, 230026, China}
\affiliation{Synergetic Innovation Center of Quantum Information and Quantum Physics, University of Science and Technology of China, Hefei, Anhui 230026, China}
\author{Xiaoling Cui}
\email{xlcui@iphy.ac.cn}
\affiliation{Beijing National Laboratory for Condensed Matter Physics, Institute of Physics, Chinese Academy of Sciences, Beijing 100190, China}
\author{Wei Yi}
\email{wyiz@ustc.edu.cn}
\affiliation{Key Laboratory of Quantum Information, University of Science and Technology of China, CAS, Hefei, Anhui, 230026, China}
\affiliation{Synergetic Innovation Center of Quantum Information and Quantum Physics, University of Science and Technology of China, Hefei, Anhui 230026, China}

\begin{abstract}
We report the existence of a universal trimer state induced by an impurity interacting with a two-component spin-orbit coupled Fermi gas in two dimensions. In the zero-density limit with a vanishing Fermi sea, the trimer is stabilized by the symmetry of the single-particle spectrum under spin-orbit coupling, and is therefore {\it universal} against the short-range details of the interaction potential. When the Fermi energy increases, we show that the trimer is further stabilized by particle-hole fluctuations over a considerable parameter region. We map out the phase diagram consisting of trimers, dimers, and polarons, and discuss the detection of these states using radio-frequency spectroscopy. The universal trimer revealed in our work is a direct manifestation of intriguing three-body correlations emerging from a many-body environment, which, in our case, is cooperatively supported by the single-particle spectral symmetry and the collective particle-hole excitations.
\end{abstract}
\pacs{67.85.Lm, 03.75.Ss, 05.30.Fk}

\maketitle

\emph{Introduction}.--
The problem of mobile impurities in a many-body environment has generated significant research interests in various fields of physics. In ultracold atomic gases in particular, thanks to the clean environment and high tunability, systems consisting of a few impurities in the backdrop of a quantum degenerate gas serve as ideal platforms for investigating the interplay between few- and many-body physics~\cite{mitpolaron,enspolaron,grimmpolaron,kohlpolaron,polarondynamics,bosonpolaronexp}. In these systems, few-body correlations can have decisive impact on many-body properties, and thus manifest themselves more transparently in the latter. For instance, the two-body correlations in attractive and repulsive branches of spin-imbalanced Fermi gases are closely related to pairing and magnetism, respectively, of the corresponding many-body systems in the high-polarization limit~\cite{impurityreview1,impurityreview2,mag2,mag3}. Very recently, the more unusual three-body correlations have also been discovered for impurities in mass-imbalanced fermions~\cite{3body2,parishprl,parishpra,3body3,naidonpra}, as well as on top of a Fermi or Bose superfluid~\cite{3body5,3body6,bosonpolaronexp,3body7,3body8}. These findings form the building blocks in the search of intriguing quantum states in a many-body system with exotic few-body correlations.

In all these previous studies on three-body correlations~\cite{bosonpolaronexp,3body5,3body6,3body7,3body8, 3body2,parishprl,parishpra,3body3,naidonpra}, the states are either non-universal, i.e., dependent on the short-range interaction details, or not easily accessible by current experiments due to stringent requirements such as large mass ratios. A natural question is whether one can find a system with dominant three-body correlations that are universal and more accessible. In this context, a promising system is impurities immersed in a spin-orbit coupled Fermi gas. By modifying the single-particle dispersion, spin-orbit coupling (SOC) can lead to a variety of interesting quantum states, which have attracted much interest in the field of cold atoms following its experimental realizations~\cite{gauge2exp,fermisocexp1,fermisocexp2,2dsoczhang1,Wu2015,socreview1,socreview2,socreview4,socreview5,socreview6}. In particular, it has been shown that SOCs with high symmetries can favor the formation of universal trimers in three dimensional three-body systems~\cite{3dsoc3body,borromean}. Nevertheless, the stability of these SOC-induced trimers in a many-body environment still remains to be investigated.

In this work, we consider an impurity interacting spin selectively with a Fermi gas under the Rashba SOC. To simplify numerical calculation, we consider a two-dimensional Fermi gas. We further assume that the Fermi gas consists of two spin components that are non-interacting with each other. A similar setup with an equal Rashba and Dresselhaus mixed SOC has been studied previously, which offers a novel Fulde-Ferrell pairing mechanism~\cite{zhouprl}. Here, we show that in the zero-density limit of the Fermi gas, a universal trimer is stabilized over a wide interaction range by the single-particle symmetry under the Rashba SOC. This is analogous to the universal Borromean state in three dimensions~\cite{borromean}. By slightly increasing the fermion density from zero, we find that the universal trimer is further stabilized such that it exists in a broader parameter regime intervening between the polaron and the dimer regimes. We systematically analyze the impact of the Fermi sea, and show that, while the effects of Pauli blocking and particle-hole excitations compete with one another, the latter particularly favors trimer formation. We map out the phase diagrams with varying interaction strength, Fermi energy, and mass ratio, and discuss the detection of different states using radio-frequency (r.f.) spectroscopy in cold-atoms experiments. By revealing stable universal trimers in impurity systems, our work offers valuable insights that are helpful to the on-going effort in bridging few- and many-body physics with cold atoms.

\emph{Model}.--
The Hamiltonian of the system can be written as
\begin{align}
H&=\sum_{{\bf k}\sigma=\uparrow,\downarrow}\frac{{\hbar^2\bf k}^2}{2m_a} a^{\dag}_{{\bf k}\sigma}a_{{\bf k}\sigma}+\sum_{\bf k}\left[\alpha (k_x-ik_y ) a^{\dag}_{{\bf k}\uparrow}a_{{\bf k}\downarrow} + H.c. \right]\nonumber\\
&+\sum_{\bf k}\frac{{\hbar^2\bf k}^2}{2m_b} b^{\dag}_{\bf k}b_{\bf k}+\frac{U}{V}\sum_{{\bf k,k',q}}a^{\dag}_{{\bf k}\uparrow}b^{\dag}_{\bf q-k}b_{\bf q-k'}a_{{\bf k'}\uparrow}, \label{H}
\end{align}
where $b_{\bf k}$ annihilates an impurity atom with mass $m_b$, and $a_{\cp k\sigma}$ annihilates an atom in the Fermi sea with spin $\sigma$ and mass $m_a$; $\alpha$ is the strength of Rashba SOC; $U$ is the bare interaction between the impurity and spin-up atoms, which can be renormalized as $1/U=-(1/V)\sum_{\bf k} 1/(E_b+(1+\eta) \hbar^2k^2/m_a)$; $V$ and $E_b$ are respectively the quantization area and two-body binding energy in 2D. Finally we denote the mass ratio as $\eta=m_a/m_b$.

Diagonalizing the single-particle Hamiltonian of a spin-orbit coupled fermion, we have the creation operators of the helicity states $a^{\dag}_{{\bf k}\lambda}=\sum_{\sigma} \gamma^{\sigma}_{{\bf k}\lambda} a^{\dag}_{{\bf k}\sigma}$, with the eigen energies $\xi_{\cp k\pm}=\hbar^2/2m_a(|\cp k|\pm k_0)^2+E_{th}$. Here, $\lambda=\pm,\ \gamma^{\uparrow}_{{\bf k}\pm}=1/\sqrt{2},\ \gamma^{\downarrow}_{{\bf k}\pm}= \pm e^{ -i\phi_{k}}/\sqrt{2}$, $\phi_k={\rm arg}(k_x,k_y)$, $k_0=m_a\alpha/\hbar^2$, and the ground-state threshold energy $E_{th}=-m_a \alpha^2/2\hbar^2$. The single-particle ground state under the Rashba SOC has a U(1) degeneracy along a circle in momentum space with a radius $k_0$. In the following, we will use $k_0$ and $E_0=|E_{th}|$ as the units of wave vector and energy, respectively.

\emph{Trimer and dimer in the few-body sector}.--
We first examine a three-body system with an impurity and two fermions in the helicity states. The wave function of a trimer can be written as
\begin{align}
|\psi_{\cp Q}^{(3)}\rangle=\sum_{\cp k\lambda\cp k'\beta}\psi^{\lambda\beta}_{\cp k\cp k'}(\cp Q)b^{\dag}_{\cp Q-\cp k-\cp k'}a^{\dag}_{\cp k\lambda}a^{\dag}_{\cp k'\beta}|0\rangle,
\end{align}
where $\cp Q$ is the center-of-mass (CoM) momentum of the trimer. The trimer energy $E_3$, relative to the three-body threshold $2E_{th}$, can be calculated by solving the Schr\"odinger's equation $H|\psi_{\cp Q}^{(3)}\rangle=(E_3+2E_{th})|\psi_{\cp Q}^{(3)}\rangle$. Similarly, we can evaluate the dimer energy, relative to the two-body threshold $E_{th}$, through its wave function
\begin{align}
|\psi_{\cp Q}^{(2)}\rangle=\sum_{\cp k\lambda}\psi^{\lambda}_{\cp k}(\cp Q)b^{\dag}_{\cp Q-\cp k}a^{\dag}_{\cp k\lambda}|0\rangle.
\end{align}

\begin{figure}[tbp]
\includegraphics[width=8cm]{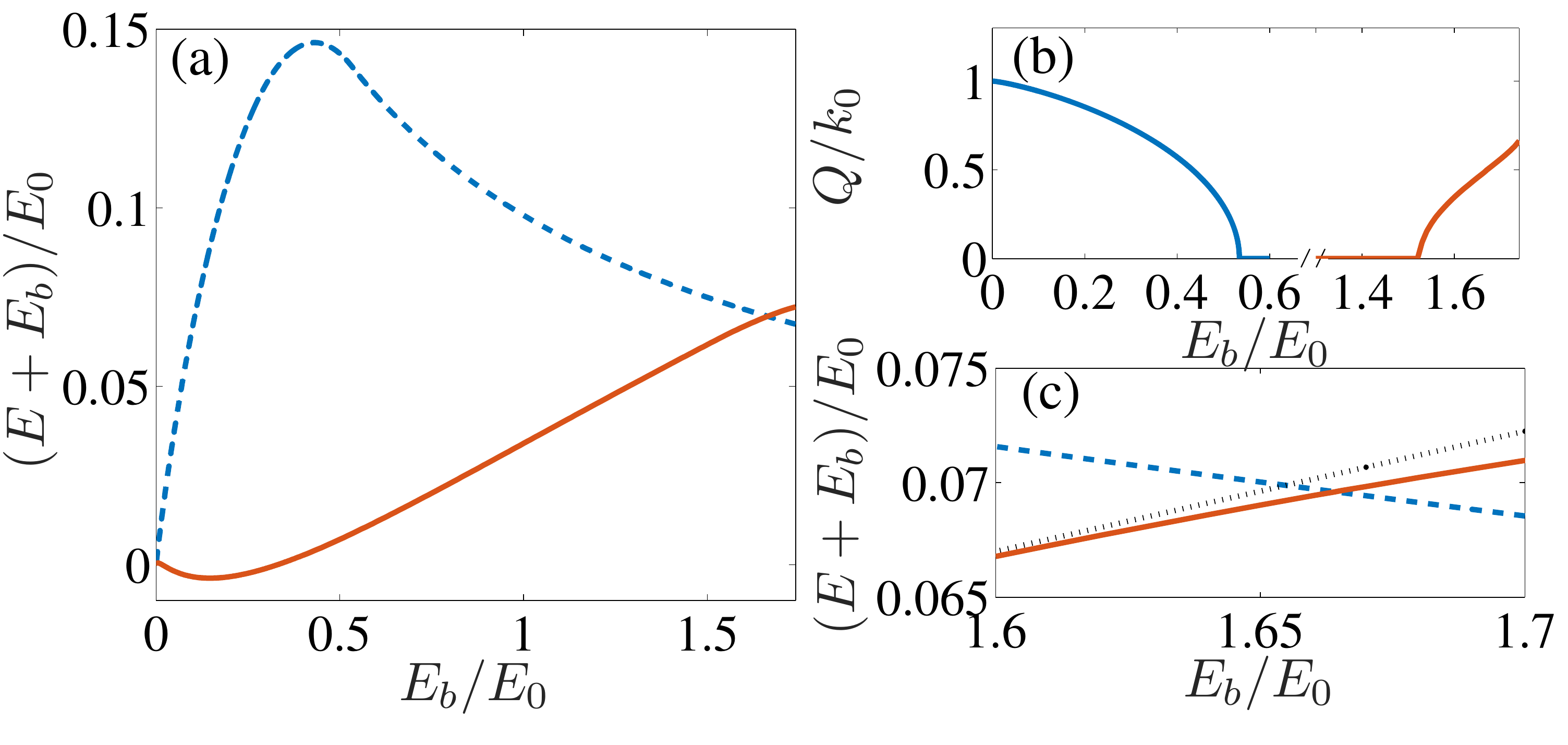}
\caption{(Color online) (a) Trimer (red solid) and dimer (blue dashed) energies in a three-body system as functions of $E_b$. (b) Total momenta of dimer (blue) and trimer (red) as functions of $E_b$. (c) Trimer and dimer energies near the transition point. The black-dotted line shows the energy of the zero-momentum trimer, which is very close to that of the ground-state trimer with a finite momentum. }
\label{fig:fig1}
\end{figure}

We compare trimer and dimer energies for the equal mass case $\eta=1$ in Fig.~\ref{fig:fig1}. Apparently, the trimer is stable over a considerable interaction range with $E_b\in [0,1.66]E_0$. Similar to the three-dimensional case, the stabilization of the trimer here is directly related to the SOC-induced U(1) degeneracy of the single-particle ground state, which particularly favors three-body, rather than two-body, scattering in the low-energy subspace~\cite{borromean}. Remarkably, the trimer energy is universally determined by the physical parameters ($E_b$, $\alpha$ and $\eta$), and is robust against short-range details of the interaction potential. For such a universal trimer, the momentum distribution of the spin-orbit coupled fermions has the largest weight on the U(1)-degenerate ring with a radius $k=k_0$~\cite{supplement}. With increasing $E_b$, the SOC physics becomes less dominant compared to interaction effects, and the dimer eventually takes over the trimer as the ground state for $E_b\ge 1.66 E_0$ (Fig.~\ref{fig:fig1}(a)).

In 2D, both the trimer and the dimer can acquire a finite CoM momentum $\cp Q$ (Fig.~\ref{fig:fig1}(b)). However, as the energy difference between $|\psi^{(3)}_{Q=0}\rangle$ and the ground-state $|\psi^{(3)}_{\cp Q}\rangle$ is typically quite small up to the trimer-dimer transition (Fig.~\ref{fig:fig1}(c)), we estimate the trimer-dimer transition using a zero-momentum trimer. For the following discussions in the presence of spin-orbit coupled Fermi sea, we will only consider the dressed trimer in the $Q=0$ sector, which would yield a lower-bound estimation of the stability region of the dressed trimer.

\emph{Trimer, dimer and polaron in the presence of a Fermi sea}.--
When the impurity is immersed in a spin-orbit coupled Fermi sea, we consider the following ansatz for the trimer dressed by particle-hole excitations~\cite{parishprl,parishpra}
\begin{align}
|T_0\rangle&=\sum_{\cp k\lambda\cp k'\beta}{}^{'}\phi^{\lambda\beta}_{\cp k\cp k'}b^{\dag}_{-\cp k-\cp k'}a^{\dag}_{\cp k\lambda}a^{\dag}_{\cp k'\beta}|{\rm FS}\rangle_{N-2}\nonumber\\
&+\sum_{\substack{\cp k\lambda \cp k'\beta \\ \cp k''\gamma \cp q\nu}}{}^{'} \phi^{\lambda\beta\gamma\nu}_{\cp k\cp k'\cp k''\cp q}b^{\dag}_{\cp q-\cp k-\cp k'-\cp k''}a^{\dag}_{\cp k\lambda}a^{\dag}_{\cp k'\beta}a^{\dag}_{\cp k''\gamma}a_{\cp q\nu}|{\rm FS}\rangle_{N-2},\label{eqn:dressedtrimer}
\end{align}
where $|{\rm FS}\rangle_N$ represents a spin-orbit coupled Fermi sea with $N$ atoms. For the summations above, we have $\xi_{\cp k\lambda},\xi_{\cp k'\beta},\xi_{\cp k''\gamma}>E_F$, and $\xi_{\cp q\nu}<E_F$, where $E_F$ is the Fermi energy of the spin-orbit coupled fermions. The first term on the right-hand side of Eq.~(\ref{eqn:dressedtrimer}) represents a bare trimer on top of the Fermi sea, and the second term accounts for contributions from a single pair of particle-hole fluctuations. We expect the energy of the dressed trimer could be estimated fairly accurately at this level~\cite{supplement}. Moreover, our ansatz recovers the exact few-body wave function when the fermion density is sent to zero.

Minimizing the energy functional $\langle T_0| H-E_T-2E_F|T_0\rangle$, we get a set of coupled equations for $F^{\lambda}_{\cp k}$ and $G^{\lambda\beta\nu}_{\cp k\cp k'\cp q}$, where $F^{\lambda}_{\cp k}=\sum_{\cp k'\beta}\psi^{\lambda\beta}_{\cp k\cp k'}$, and $G^{\lambda\beta\nu}_{\cp k\cp k'\cp q}=\psi^{\lambda\beta}_{\cp k\cp k'}+3\sum_{\cp k''\gamma}\psi^{\lambda\beta
\gamma\nu}_{\cp k\cp k'\cp k''\cp q}$. We then make the decomposition $F^{\lambda}_{\cp k}=\sum_{m}F_{m}^{\lambda}(k)\cos(m\phi_k)$, $G^{\lambda\beta\nu}_{\cp k\cp k'\cp q}=\sum_{m}G_m^{\lambda\beta\nu}(k,k',q,\hat{\cp k}\cdot\hat{\cp k'},\hat{\cp k}\cdot\hat{\cp q})\cos(m\phi_k)$. Due to the conservation of quantum number $m$, one can solve $E_T$ in any given $m$-sector. Note that the trimer energy $E_T$ is relative to the Fermi energy of a spin-orbit coupled Fermi sea of $N$ atoms. To further simplify the numerics, we make the approximation $|{\bf q}|=k_0$, i.e. the hole excitation occurs only on the degenerate ring of the single-particle ground state. This approximation should be particularly good for a dilute gas with a small Fermi energy.

\begin{figure}[tbp]
\includegraphics[width=8cm]{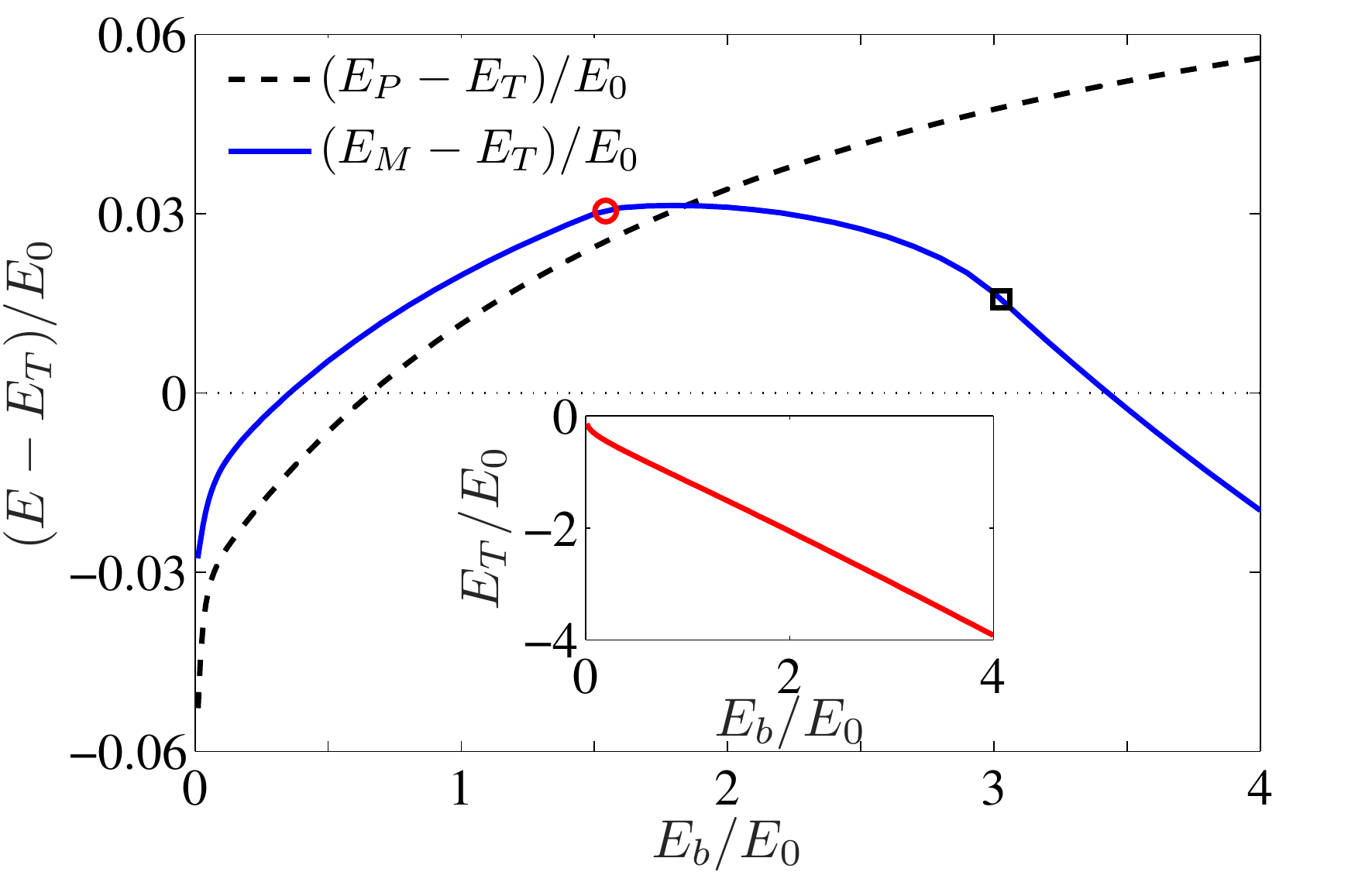}
\caption{(Color online) Energies of dressed dimer (blue solid) and polaron (black dashed) relative to the energy of dressed trimer as functions of $E_b$. For all three states, the energy reference is the total energy of a spin-orbit coupled Fermi sea of $N$ atoms. Inset: Energy of the dressed trimer. Here, the Fermi energy is $E_F=E_{th}+ 0.05 E_0$, and the mass ratio $\eta=1$. The red circle marker at $E_b\approx 1.54E_0$ indicates a first-order transition between dimer states with different CoM momentum, and the black square marker at $E_b\approx 3.03E_0$ indicates a second-order transition between $D_Q$ and $D_0$~\cite{supplement}. The CoM momentum of the polaron is always zero for the parameters that we consider.
}
\label{fig:fig2}
\end{figure}

Similar to the few-body case, as $E_b$ increases the effect of SOC becomes less important. The ground state of the system in the strong-coupling limit should be a dimer dressed by particle-hole excitations. We consider a dressed-dimer ansatz up to a single pair of particle-hole excitations~\cite{impurityreview1,impurityreview2,combescot}
\begin{align}
|D_{\cp Q}\rangle&=\sum_{\cp k\lambda}{}^{'}\phi^{\lambda}_{\cp k}(\cp Q)b^{\dag}_{\cp Q-\cp k}a^{\dag}_{\cp k\lambda}|{\rm FS}\rangle_{N-1}\nonumber\\
&+\sideset{}{'}\sum_{\cp k\lambda\cp k'\beta\cp q\nu} \phi^{\lambda\beta\nu}_{\cp k\cp k'\cp q}(\cp Q)b^{\dag}_{\cp Q-\cp k-\cp k'+\cp q}a^{\dag}_{\cp k\lambda}a^{\dag}_{\cp k'\beta}a_{\cp q\nu}|{\rm FS}\rangle_{N-1}.\label{eqn:dresseddimer}
\end{align}
For a fermionic impurity in the thermodynamic limit, $|D_{\cp Q}\rangle$ corresponds to a particle-hole-dressed Bardeen-Cooper-Schieffer pairing state in the large polarization limit for $Q=0$, and a Fulde-Ferrell-Larkin-Ovchinnikov state if $\cp Q$ is finite.

When the interaction is weak, one should also consider the possibility of a polaron~\cite{impurityreview1,impurityreview2,combescot}
\begin{align}
\left|P_{\cp Q}\right\rangle=\Big(\phi_{\cp Q} b^{\dag}_{\cp Q} +\sum_{\cp k\lambda\cp q\nu}{}^{'} \phi^{\lambda\nu}_{\cp k\cp q}(\cp Q)b^{\dag}_{\cp Q+\cp q-\cp k}a^{\dag}_{\cp k\lambda}a_{\cp q\nu}\Big) \left|{\rm FS}\right\rangle_N.\label{eqn:polaron}
\end{align}
In the thermodynamic limit, the polaron corresponds to a particle-hole-dressed normal state.

In Fig.~\ref{fig:fig2}, we show energies of the dressed trimer, the dressed dimer, and the polaron as functions of $E_b$ at a typical Fermi energy $E_F=E_{th}+ 0.05 E_0$. Consistent with our expectation, the ground state is the polaron and the dressed dimer, respectively, in the weak- and the strong-coupling limit. Importantly, the dressed trimer is stable over a fairly large parameter region with $E_b\in[0.66,3.43]E_0$, which is considerably broadened as compared to the few-body case (Fig.~\ref{fig:fig1}(a)).

\begin{figure}[tbp]
\includegraphics[width=8cm]{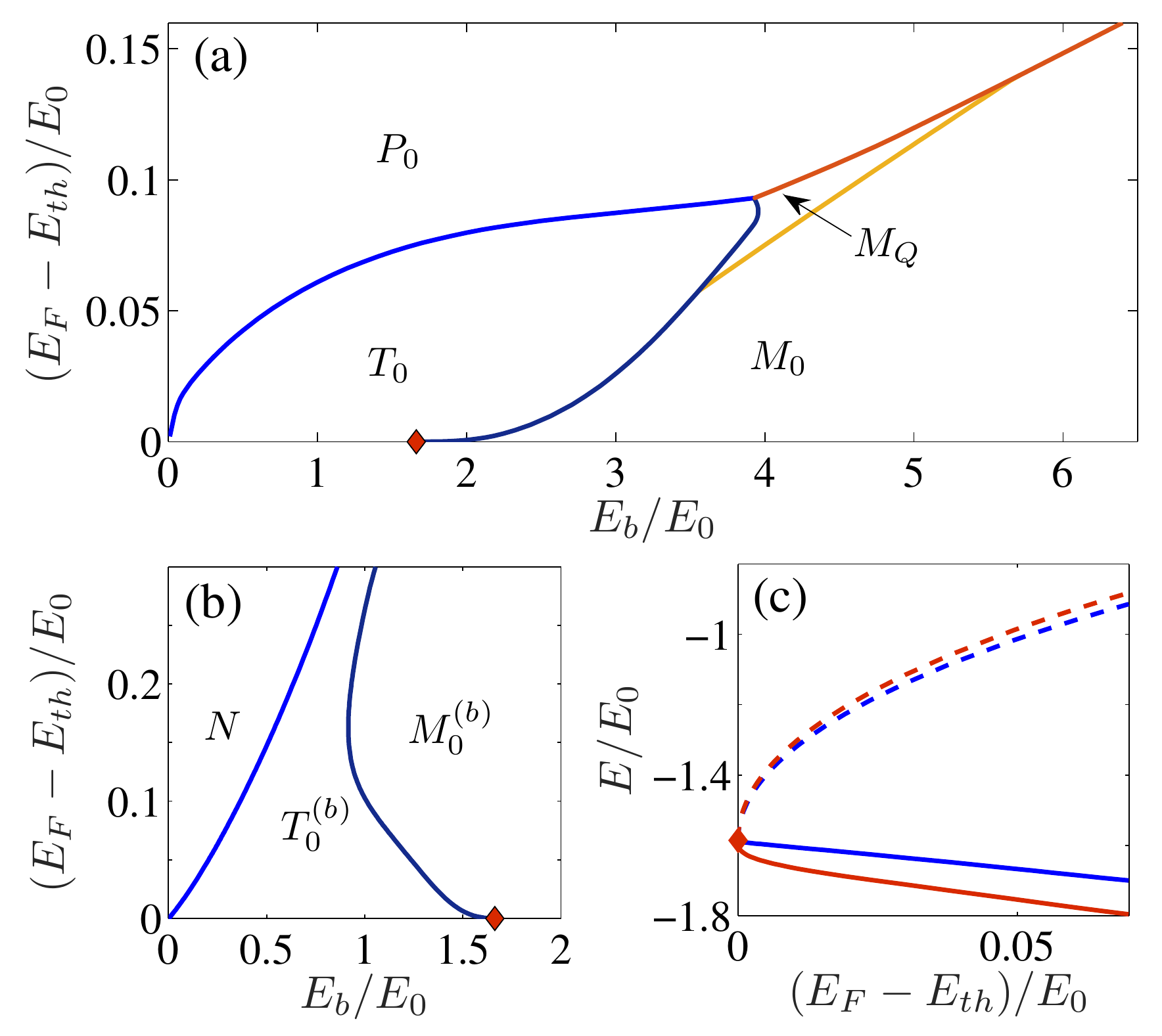}
\caption{(Color online) (a) Phase boundaries between the dressed trimer state ($T_0$), the dressed dimer with zero ($D_0$) and finite ($D_Q$) total momentum, and the polaron with zero total momentum ($P_0$). The red diamond indicates the few-body transition point between $|\psi^{(3)}_{0}\rangle$ and $|\psi^{(2)}_0\rangle$ at $E_b\approx 1.65E_0$. (b) Phase boundaries between the bare trimer ($T^{(b)}$), the bare dimer ($D^{(b)}$), and the normal state ($N$). (c) Energies of trimer (red) and dimer (blue) as functions of $E_F$ at $E_b=1.65E_0$. The solid (dashed) lines are energies of dressed (bare) states. Here the mass ratio $\eta=1$.}
\label{fig:fig3}
\end{figure}

\emph{Phase diagrams and effects of Fermi sea}.--
In Fig.~\ref{fig:fig3}(a), we map out the phase diagram involving all different states on the $E_F$--$E_b$ plane. Remarkably, we find that, as $E_F$ increases from $E_{th}$ in the low-density limit, the dressed trimer ($T_0$) becomes more stable against the dressed dimer ($D_0$). As $E_F$ further increases, $T_0$ gives way to the polaron ($P_0$). Furthermore, a dimer with a finite CoM momentum ($D_Q$) emerges in a narrow region surrounded by $P_0$, $D_0$ and $T_0$.

\begin{figure}[tbp]
\includegraphics[width=8cm]{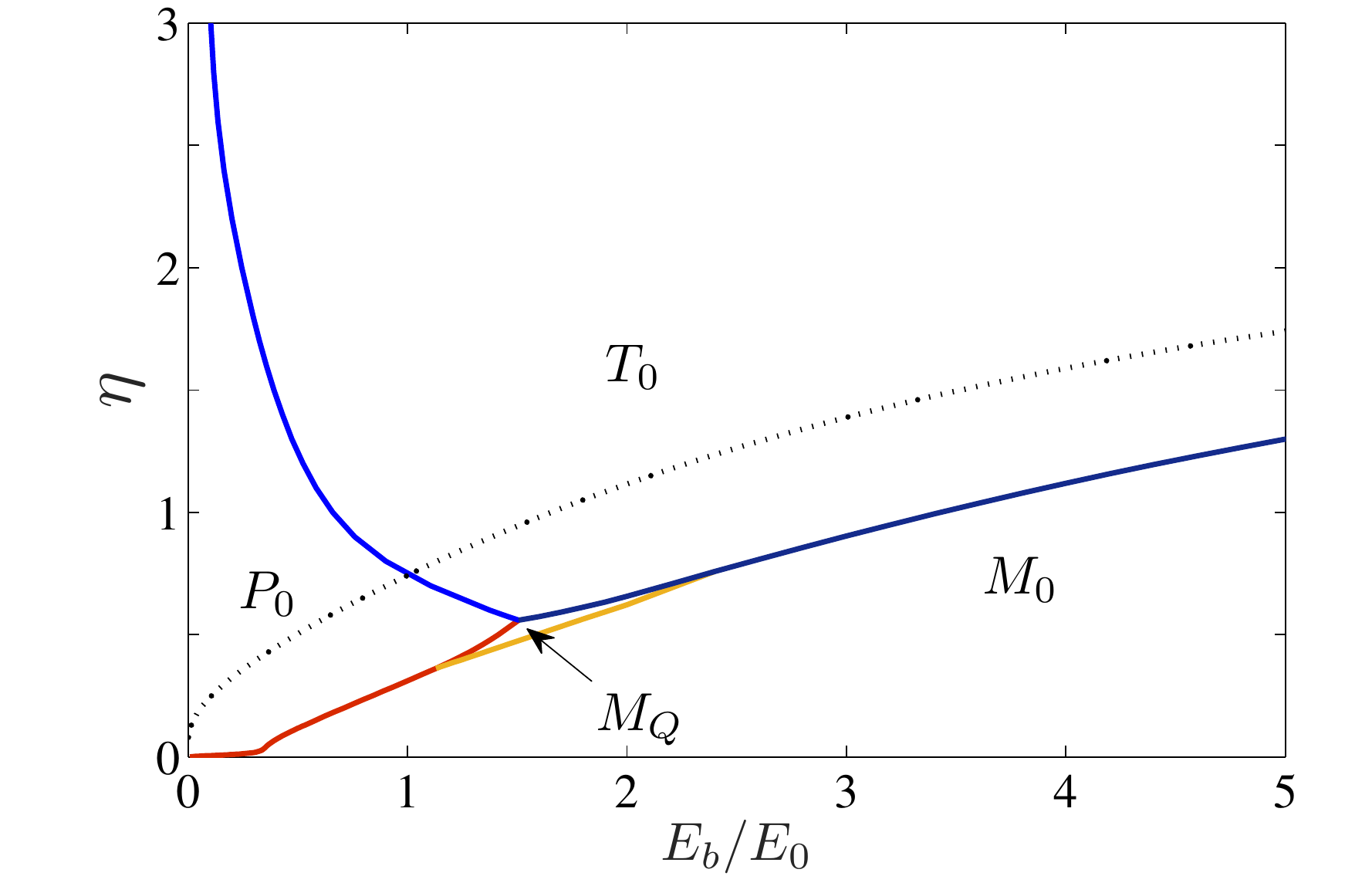}
\caption{(Color online) Phase diagram on the $\eta$--$E_b$ plane with a fixed $E_F=E_{th}+ 0.05 E_0$. The black dots indicates the boundary between trimer (toward the upper left) and dimer (toward the lower right) in the few-body sector. }
\label{fig:fig4}
\end{figure}

The enhanced stability of trimer against dimer as shown in Fig.~\ref{fig:fig3}(a) can be understood by examining two key effects caused by the presence of Fermi-sea atoms: (i) the Pauli-blocking, which prohibits scattering within the Fermi sea; and (ii) particle-hole fluctuations, which lead to excitations out of the Fermi sea. The impact of (i) can be analyzed by keeping only the first terms in Eqs.~(\ref{eqn:dressedtrimer},\ref{eqn:dresseddimer},\ref{eqn:polaron}) and studying the phase diagram of the resulting bare states. In Fig.~\ref{fig:fig3}(b), we see that an increasing $E_F$ would favor bare dimers rather than bare trimers in the low-density limit. This behavior can be traced back to the SOC-induced trimer formation, where trimers are facilitated by the U(1) spectral symmetry of the single-particle ground states. When these states are blocked by the Fermi sea, the trimer would become unstable. In contrast, when SOC is absent, Pauli blocking would destabilize dimer more than trimer~\cite{parishprl,parishpra}. This is because the three-body scattering, which has much larger phase space than the two-body scattering, is affected less by the effect of (i). The special role of Pauli blocking in our system further underscores the uniqueness of the SOC-induced trimer formation.

From the phase boundaries in Fig.~\ref{fig:fig3}(a) and (b), it is apparent that effect (ii) plays a decisive role in stabilizing the dressed trimer. To see this explicitly, we plot the energies of dimers and trimers with increasing $E_F$ at the trimer-dimer transition in the zero-density limit (Fig.~\ref{fig:fig3}(c)). For the bare states with only effect (i), both the dimer and the trimer energies would increase with $E_F$, while the bare trimer is higher in energy. However, when effect (ii) is included, both energies would decrease with increasing $E_F$, while the dressed trimer is lower in energy. An intuitive picture is that by involving states below the Fermi sea into the scattering process, particle-hole fluctuations partially recover the lost symmetries in the low-energy subspace, which are crucially important for the trimer formation.

So far, we have only considered the equal mass case $\eta=1$. When $\eta$ increases, i.e., when the impurity becomes lighter, the trimer would become more stable~\cite{parishprl,zhouprl}. In Fig.~\ref{fig:fig4}, we show the phase diagram on the $\eta$--$E_b$ plane at a fixed low Fermi energy $E_F=E_{th}+0.05E_0$. In this case, the trimer already emerges as the ground state of the system when $\eta$ is as small as $0.5$. From the trimer-dimer phase boundaries in the few-body sector and in the presence of a Fermi sea (see Fig.~\ref{fig:fig4}), we see that the presence of a Fermi sea further stabilizes the universal trimer.

\begin{figure}[tbp]
\includegraphics[width=9cm]{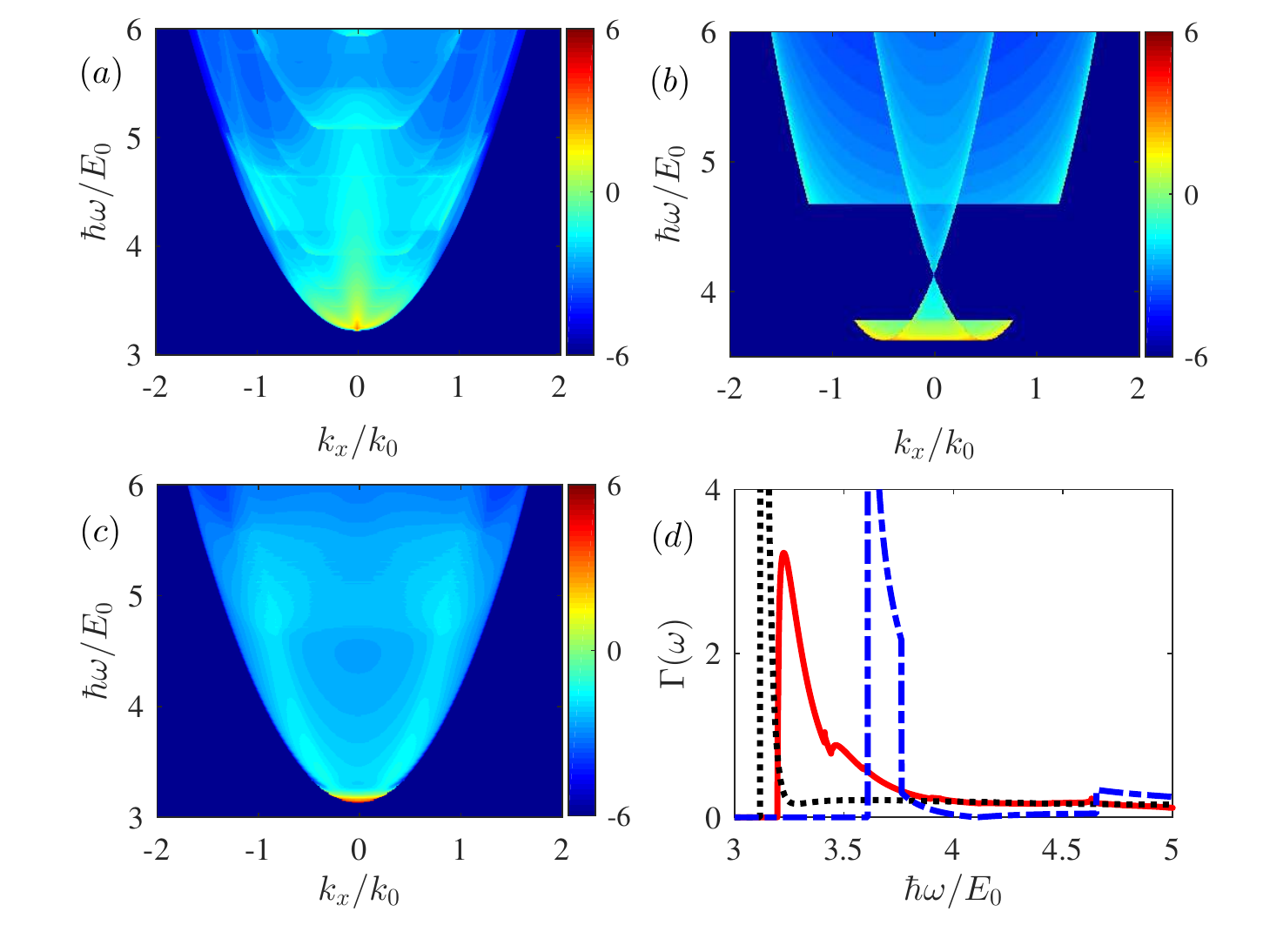}
\caption{(Color online) Contour plot of momentum-resolved r.f. spectra for (a) the dressed trimer, (b) the dressed dimer, and (c) the polaron states. The intensity is shown in the logarithmic scale. (d) The total r.f. spectra for the dressed trimer (red solid), the dressed dimer (blue dash-dotted), and the polaron (black dotted) states. We have taken $E_F= E_{\rm th}+0.05E_0$ and $\eta=1$. In the polaron spectra, we have assumed a momentum-dependent Gaussian broadening~\cite{supplement}, such that the contribution from the impurity residue, which carries most weight of the spectra, becomes visible. For all panels, we have $E_b=3.2E_0$.}
\label{fig:fig5}
\end{figure}

\emph{Detection}.--
The various states discussed here can be detected using the r.f. spectroscopy. By applying the r.f. field $H_{\rm rf}=V_0\sum_{\cp k} (c^{\dag}_{\cp k}b_{\cp k}+H.c.)$ to excite the impurity atom to a bystander state (with creation operator $c^{\dag}_{\cp k}$), one can obtain crucial information on the binding energy, the momentum distribution and the low-energy excitations of the impurity state. The total r.f. spectrum can be evaluated as $\Gamma(\omega)=2\pi/\hbar\sum_f |\langle f|H_{\rm rf}|i\rangle|^2\delta(\hbar\omega+E_i-E_f)$, where $i$ ($f$) labels the initial (final) state.
In the momentum-resolved r.f. spectroscopy, the population transfer at a given momentum $k_x$ can be experimentally probed, which can be calculated by leaving out the integration over $k_x$ in the summation of final states~\cite{supplement}.

In Fig.~\ref{fig:fig5}, we show both the total and the momentum-resolved r.f. spectra for different states. While the large impurity residue in the polaron leads to a pronounced peak at the threshold (Fig.~\ref{fig:fig5}(c)(d)), the dimer exhibits unique sharp edges in the momentum-resolved spectra (Fig.~\ref{fig:fig5}(b)) and a double-peak structure in the total spectra (Fig.~\ref{fig:fig5}(d)). These are due to the ring-topology of the Fermi sea, and consequently, the existence of two Fermi momenta~\cite{supplement}. While similar in the overall profile as the polaron spectra, the trimer spectra have less pronounced peaks and richer fine structures. These distinctive signatures should allow us to differentiate the impurity states in cold-atoms experiments. Finally, it would also be interesting to perform r.f. spectroscopy on fermions under SOC, where the impact of three-body correlations on the Fermi sea may be probed.

\emph{Summary}.--
Our results demonstrate the robustness of SOC-induced universal trimers in the many-body setting, and reveals the crucial role of particle-hole fluctuations in stabilizing these trimers. Given the stabilization mechanisms, we expect that similar universal trimers should also appear when the impurity is embedded in a three dimensional Fermi gas with a highly symmetric SOC. In practice, the highly symmetric SOC required for the trimer formation can be realized, for example, following the cyclic scheme proposed in Ref.~\cite{ian2dsoc}, or using the experimental setup in Ref.~\cite{Wu2015}. Furthermore, the existence of stable universal trimers in a many-body system opens up avenues for the study of few to many crossover physics in mesoscopic cold-atoms systems. We hope our work can stimulate more explorations of intriguing many-body phases with exotic few-body correlations.

\emph{Acknowledgments}.--
This work is supported by the National Key R\&D Program (Grant No. 2016YFA0301700), the National Natural Science Foundation of China (Grant Nos. 11374177, 11374283, 11522545, 11534014), and the programs of Chinese Academy of Sciences. W. Y. acknowledges support from the ``Strategic Priority Research Program(B)'' of the Chinese Academy of Sciences, Grant No. XDB01030200.

\newpage
\begin{widetext}
\appendix
\section{Supplemental Materials}

In this Supplemental Materials, we provide details on the solution of various states, the expression for the radio-frequency spectra, and the applicability of the ansatz wave functions for the dressed states.

\subsection{\textbf{Few-body sector}}\label{supp:fewbody}

In this section, we present the derivation of equations for the two-body and three-body bound states. The wave function for the two-body bound state can be written as
\begin{align}
|\psi^{(2)}_{\cp Q}\rangle=\sum_{\cp k\lambda}\psi^{\lambda}_{\cp k}(\cp Q)b^{\dag}_{\cp Q-\cp k}a^{\dag}_{\cp k\lambda}|0\rangle.
\end{align}
From the Schr\"odinger's equation, we have the closed equation for the two-body bound state energy $E_2$
\begin{align}
\frac{2}{U}=\sum_{\cp k\lambda}\frac{1}{E_2+E_{th}-\epsilon^b_{\cp Q-\cp k}-\xi_{\cp k\lambda}},
\end{align}
where $E_2$ is relative to the two-body threshold energy $E_{th}=-m_a\alpha^2/(2\hbar^2)$. We also have $\epsilon^b_{\cp k}=\hbar^2 k^2/2m_b$, and $\xi_{\cp k\pm}=\hbar^2/2m_a(k\pm k_0)^2+E_{th}$.

The two-body wave function can be derived as
\begin{eqnarray}
\psi^{\lambda}_{\cp k}(\cp Q)\ &\propto& \frac{1}{E_2+E_{th}-\epsilon^b_{\cp Q-\cp k}-\xi_{\cp k\lambda}}.
\end{eqnarray}

For the three-body bound state, we have
\begin{align}
|\psi^{(3)}_{\cp Q}\rangle=\sum_{\cp k\lambda\cp k'\beta}\psi^{\lambda\beta}_{\cp k\cp k'}(\cp Q)b^{\dag}_{\cp Q-\cp k-\cp k'}a^{\dag}_{\cp k\lambda}a^{\dag}_{\cp k'\beta}|0\rangle.
\end{align}
And the Schr\"odinger's equation for the three-body bound state energy $E_3$ gives
\begin{align}
\frac{2}{U}F^{\lambda}_{\cp k}=\sum_{\cp k'\beta}\frac{F^{\lambda}_{\cp k}-F^{\beta}_{\cp k'}}{E_3+2E_{th}-\epsilon^b_{\cp Q-\cp k-\cp k'}-\xi_{\cp k\lambda}-\xi_{\cp k'\beta}}, \label{eqn:E3}
\end{align}
where
\begin{align}
F^{\lambda}_{\cp k}=\sum_{\cp k'\beta}\psi^{\lambda\beta}_{\cp k\cp k'}(\cp Q).
\end{align}
The bound-state energy $E_3$, which is relative to the three-body threshold $2E_{th}$, can be solved by requiring a vanishing determinant for the coefficient matrix of Eq.~\ref{eqn:E3}.

Similarly, it is straightforward to derive
\begin{eqnarray}
\psi^{\lambda\beta}_{\cp k\cp k'}(\cp Q) &\propto& \frac{F^{\lambda}_{\cp k}-F^{\beta}_{\cp k'}}{E_3+2E_{th}-\epsilon^b_{\cp Q-\cp k-\cp k'}-\xi_{\cp k\lambda}-\xi_{\cp k'\beta}}.
\end{eqnarray}

In Fig.~\ref{fig:supp1}, we show the typical momentum distribution of the wave functions of various few-body bound states. For dimers, the largest weight of the wave function is located on a circle with a radius smaller than $k_0$; while for trimers, regardless of their CoM momenta, the largest weight of the wave function is on the U(1)-degenerate ring of the single-particle ground state with a radius $k_0$.

\begin{figure*}[tbp]
\includegraphics[width=12cm]{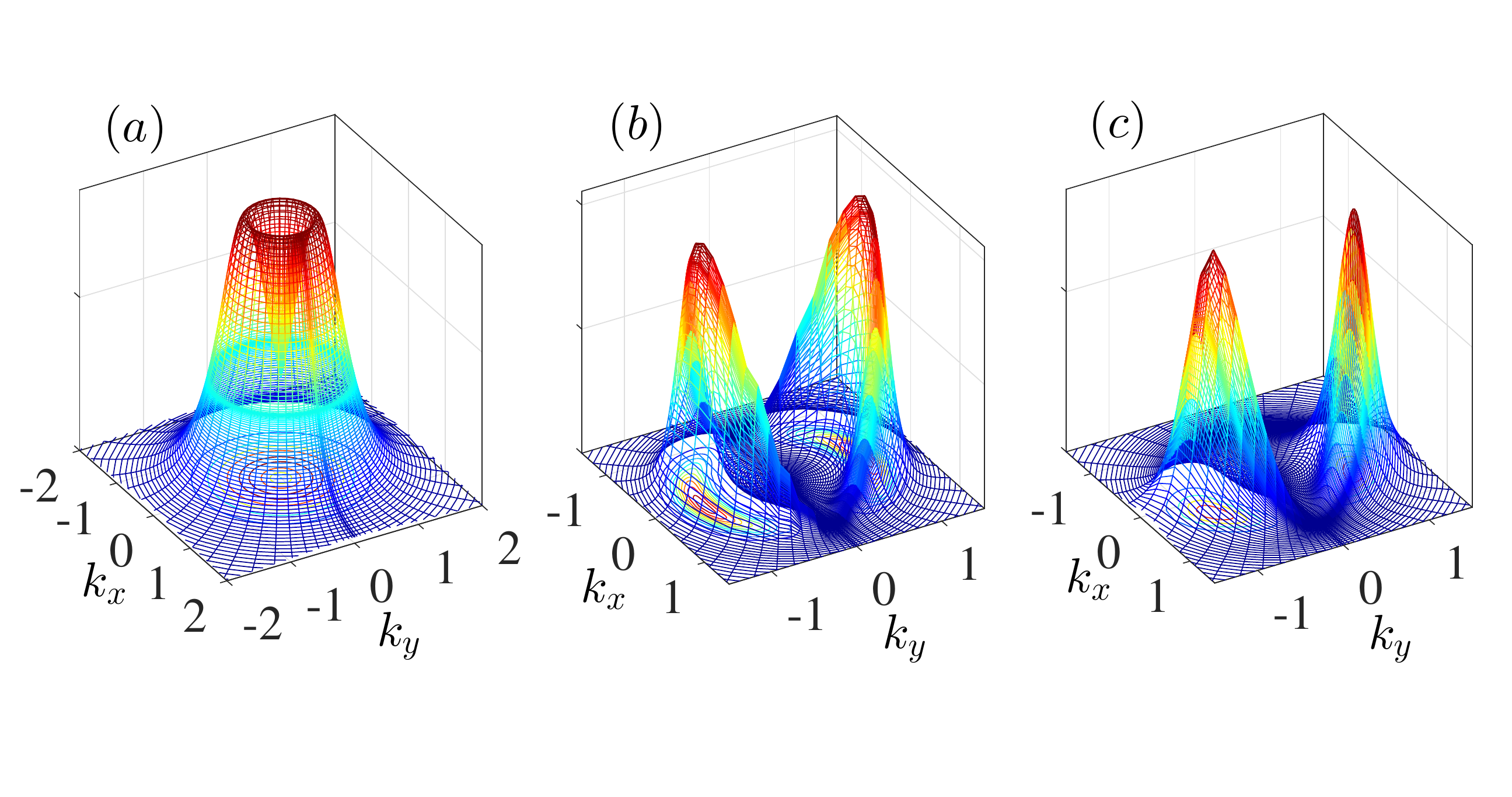}
\caption{(Color online) Momentum-space distribution of fermions in the two-body and three-body bound states. (a) Dimer with zero CoM momentum $|\psi^-_{\cp k}(0)|^2$; (b) Trimer with zero CoM momentum $|\psi^{--}_{\cp k,-\cp k}(0)|^2$; (c) Trimer with finite CoM momentum $|\psi^{--}_{\cp k,2\cp Q-\cp k}(\cp Q)|^2$. }
\label{fig:supp1}
\end{figure*}

\subsection{\textbf{Bare dimer and bare trimer}}\label{supp:barestates}

In the presence of a spin-orbit coupled Fermi sea, the simplest ansatz wave functions for the dimer and trimer states are the so-called bare dimer \begin{align}
|D^{(b)}_{\cp Q}\rangle=\sideset{}{'}\sum_{\cp k\lambda}\varphi^{\lambda}_{\cp k}(\cp Q)b^{\dag}_{\cp Q-\cp k}a^{\dag}_{\cp k\lambda}|{\rm FS}\rangle_{N-1},
\end{align}
and the bare trimer
\begin{align}
|T^{(b)}_{\cp Q}\rangle=\sideset{}{'}\sum_{\cp k\lambda\cp k'\beta}\varphi^{\lambda\beta}_{\cp k\cp k'}(\cp Q)b^{\dag}_{\cp Q-\cp k-\cp k'}a^{\dag}_{\cp k\lambda}a^{\dag}_{\cp k'\beta}|{\rm FS}\rangle_{N-2}.
\end{align}
For these bare states, we may derive the equations of their wave functions by minimizing the energy funtional $\langle H-E\rangle$. Formally, the closed/coefficient equations are the same as those in the few-body sector, except that the summation over $\cp k$ and $\cp k'$ only include states outside the Fermi sea, as those within the Fermi sea are Pauli blocked. Note that, to facilitate comparison between the energies of trimer and dimer, we fix the zero energy reference at the Fermi energy of a spin-orbit coupled Fermi sea of $N$ atoms. This convention applies to the dressed states as well.

\subsection{\textbf{Dressed trimer and dressed dimer}}\label{supp:dressedstates}

To account for contributions from particle-hole fluctuations induced by the impurity-fermion interaction, which are important in two dimensions, we consider the dressed trimer and dress dimer. For the dressed trimer, the ansatz wave function is
\begin{align}
|T_0\rangle=\sideset{}{'}\sum_{\cp k\lambda\cp k'\beta}\phi^{\lambda\beta}_{\cp k\cp k'}b^{\dag}_{-\cp k-\cp k'}a^{\dag}_{\cp k\lambda}a^{\dag}_{\cp k'\beta}|{\rm FS}\rangle_{N-2}+\sideset{}{'}\sum_{\substack{\cp k\lambda \cp k'\beta \\ \cp k''\gamma \cp q\nu}} \phi^{\lambda\beta\gamma\nu}_{\cp k\cp k'\cp k''\cp q}b^{\dag}_{\cp q-\cp k-\cp k'-\cp k''}a^{\dag}_{\cp k\lambda}a^{\dag}_{\cp k'\beta}a^{\dag}_{\cp k''\gamma}a_{\cp q\nu}|{\rm FS}\rangle_{N-2}.
\end{align}
Applying the variational approach, we arrive at a set of coupled equations
\begin{align}
&(\frac{2}{U}-\sideset{}{'}\sum_{\cp k'\beta}\frac{1}{A^{\lambda\beta}_{\cp k\cp k'}})F^{\lambda}_{\cp k}=\sideset{}{'}\sum_{\cp k'\beta}\frac{\sideset{}{'}\sum_{\cp q\nu}G^{\lambda\beta\nu}_{\cp k\cp k'\cp q}-F^{\beta}_{\cp k'}}{A^{\lambda\beta}_{\cp k\cp k'}},\\
&(\frac{2}{U}-\sideset{}{'}\sum_{\cp k''\gamma}\frac{1}{A^{\lambda\beta\gamma\nu}_{\cp k\cp k'\cp k''\cp q}})G^{\lambda\beta\nu}_{\cp k\cp k'\cp q}=\sideset{}{'}\sum_{\cp k''\gamma}\frac{G^{\beta\gamma\nu}_{\cp k'\cp k''\cp q}-G^{\lambda\gamma\nu}_{\cp k\cp k''\cp q}}{A^{\lambda\beta\gamma\nu}_{\cp k\cp k'\cp k''\cp q}} +\frac{F^{\lambda}_{\cp k}-F^{\beta}_{\cp k'}+\sideset{}{'}\sum_{\cp q\nu} G^{\lambda\beta\nu}_{\cp k\cp k'\cp q}}{A^{\lambda\beta}_{\cp k\cp k'}},
\end{align}
where we have defined
\begin{align}
F^{\lambda}_{\cp k}&=\sideset{}{'}\sum_{\cp k'\beta}\phi^{\lambda\beta}_{\cp k\cp k'},\\
G^{\lambda\beta\nu}_{\cp k\cp k'\cp q}&=\phi^{\lambda\beta}_{\cp k\cp k'}+3\sideset{}{'}\sum_{\cp k''\gamma}\phi^{\lambda\beta\gamma\nu}_{\cp k\cp k'\cp k''\cp q}, \label{G}\\
A^{\lambda\beta}_{\cp k\cp k'}&=E_T+2E_F-\xi_{\cp k\lambda}-\xi_{\cp k'\beta}-\epsilon^b_{-\cp k-\cp k'},\\
A^{\lambda\beta\gamma\nu}_{\cp k\cp k'\cp k''\cp q}&=E_T+2E_F-\xi_{\cp k\lambda}-\xi_{\cp k'\beta}-\xi_{\cp k''\gamma}+\xi_{\cp q\nu}-\epsilon^b_{\cp q-\cp k-\cp k'-\cp k''}.
\end{align}

As discussed in the main text, for the calculation of dressed-trimer energy, we make the assumptions that $q=k_0$, and that theCoM momentum is zero. Under these assumptions, we can further make the decomposition
\begin{align}
F^{\lambda}_{\cp k}&=\sum_{m}F^{\lambda}_{m}(k)\cos(m\phi_k)\nonumber\\
G^{\lambda\beta\nu}_{\cp k\cp k'\cp q}&=\sum_{m}G^{\lambda\beta\nu}_{m}(k,k',q,\hat{\cp k}\cdot\hat{\cp k'},\hat{\cp k}\cdot\hat{\cp q})\cos(m\phi_k).
\end{align}
We can then write down the coupled equations for the variables $F^{\lambda}_{m}(k)$ and $G^{\lambda\beta\nu}_{m}(k,k',q,\hat{\cp k}\cdot\hat{\cp k'},\hat{\cp k}\cdot\hat{\cp q})$. We find that the ground state of the dressed trimer lies in the $m=1$ sector.

Likewise, the ansatz wave function for the dressed dimer is
\begin{align}
|D_{\cp Q}\rangle=\sum_{\cp k\lambda}{}^{'}\phi^{\lambda}_{\cp k}(\cp Q)b^{\dag}_{\cp Q-\cp k}a^{\dag}_{\cp k\lambda}|FS\rangle_{N-1}+\sideset{}{'}\sum_{\cp k\lambda\cp k'\beta\cp q\nu} \phi^{\lambda\beta\nu}_{\cp k\cp k'\cp q}(\cp Q)b^{\dag}_{\cp Q-\cp k-\cp k'+\cp q}a^{\dag}_{\cp k\lambda}a^{\dag}_{\cp k'\beta}a_{\cp q\nu}|FS\rangle_{N-1}.
\end{align}
The Schr\"odinger's equation then gives
\begin{align}
&C+\sideset{}{'}\sum_{\cp q\nu}G^{\lambda\nu}_{\cp k\cp q}=\frac{2}{U}A^{\lambda}_{\cp k}\phi^{\lambda}_{\cp k}(\cp Q),\\
&G^{\lambda\nu}_{\cp k\cp q}-G^{\beta\nu}_{\cp k'\cp q}=\frac{4}{U}A^{\lambda\beta\nu}_{\cp k\cp k'\cp q}\phi^{\lambda\beta\nu}_{\cp k\cp k'\cp q}(\cp Q),
\end{align}
where we have defined
\begin{align}
C&=\sideset{}{'}\sum_{\cp k\lambda}\phi^{\lambda}_{\cp k}(\cp Q),\\
G^{\lambda\nu}_{\cp k\cp q}&=\phi^{\lambda}_{\cp k}(\cp Q)+2\sideset{}{'}\sum_{\cp k'\beta}\phi^{\lambda\beta\nu}_{\cp k\cp k'\cp q}(\cp Q), \label{G}\\
A^{\lambda}_{\cp k}&=E_D+E_F-\xi_{\cp k\lambda}-\epsilon^b_{\cp Q-\cp k},\\
A^{\lambda\beta\nu}_{\cp k\cp k'\cp q}&=E_D+E_F-\xi_{\cp k\lambda}-\xi_{\cp k'\beta}+\xi_{\cp q\nu}-\epsilon^b_{\cp Q-\cp k-\cp k'+\cp q}.
\end{align}
We can rearrange these into the following equations
\begin{align}
&(\frac{2}{U}-\sideset{}{'}\sum_{\cp k\lambda}\frac{1}{A^{\lambda}_{\cp k}})C=\sideset{}{'}\sum_{\cp k'\beta}\frac{\sideset{}{'}\sum_{\cp q\nu}G^{\beta\nu}_{\cp k'\cp q}}{A^{\beta}_{\cp k'}},\\
&(\frac{2}{U}-\sideset{}{'}\sum_{\cp k'\beta}\frac{1}{A^{\lambda\beta\nu}_{\cp k\cp k'\cp q}})G^{\lambda\nu}_{\cp k\cp q}=-\sideset{}{'}\sum_{\cp k'\beta}\frac{G^{\beta\nu}_{\cp k'\cp q}}{A^{\lambda\beta\nu}_{\cp k\cp k'\cp q}} +\frac{C+\sideset{}{'}\sum_{\cp q\nu} G^{\lambda\nu}_{\cp k\cp q}}{A^{\lambda}_{\cp k}}.
\end{align}
One can plug the expression of $C$ into the equation for $G_{kq}$ and solve a single matrix equation for $G^{\lambda\nu}_{\cp k\cp q}$. An interesting feature of the dressed dimer is the transitions between states with different CoM momentum as shown in Fig. 2 of the main text. At large $E_b$, the lowest-energy dimer has zero CoM momentum. As $E_b$ decreases, the dimer acquires a finite CoM momentum through a second-order transition (blue square in Fig.~\ref{fig:suppdimer}). When $E_b$ decreases further, a double-well structure emerges in the $E_b$ vs. $Q$ plot (see Fig.~\ref{fig:suppdimer}), such that the dimer with small $Q$ becomes metastable through a first-order transition (red circle in Fig.~\ref{fig:suppdimer}). As $E_b$ approaches zero, the lowest-energy dimer state has a CoM momentum $Q$ approaching $k_0$.

\begin{figure*}[tbp]
\includegraphics[width=10cm]{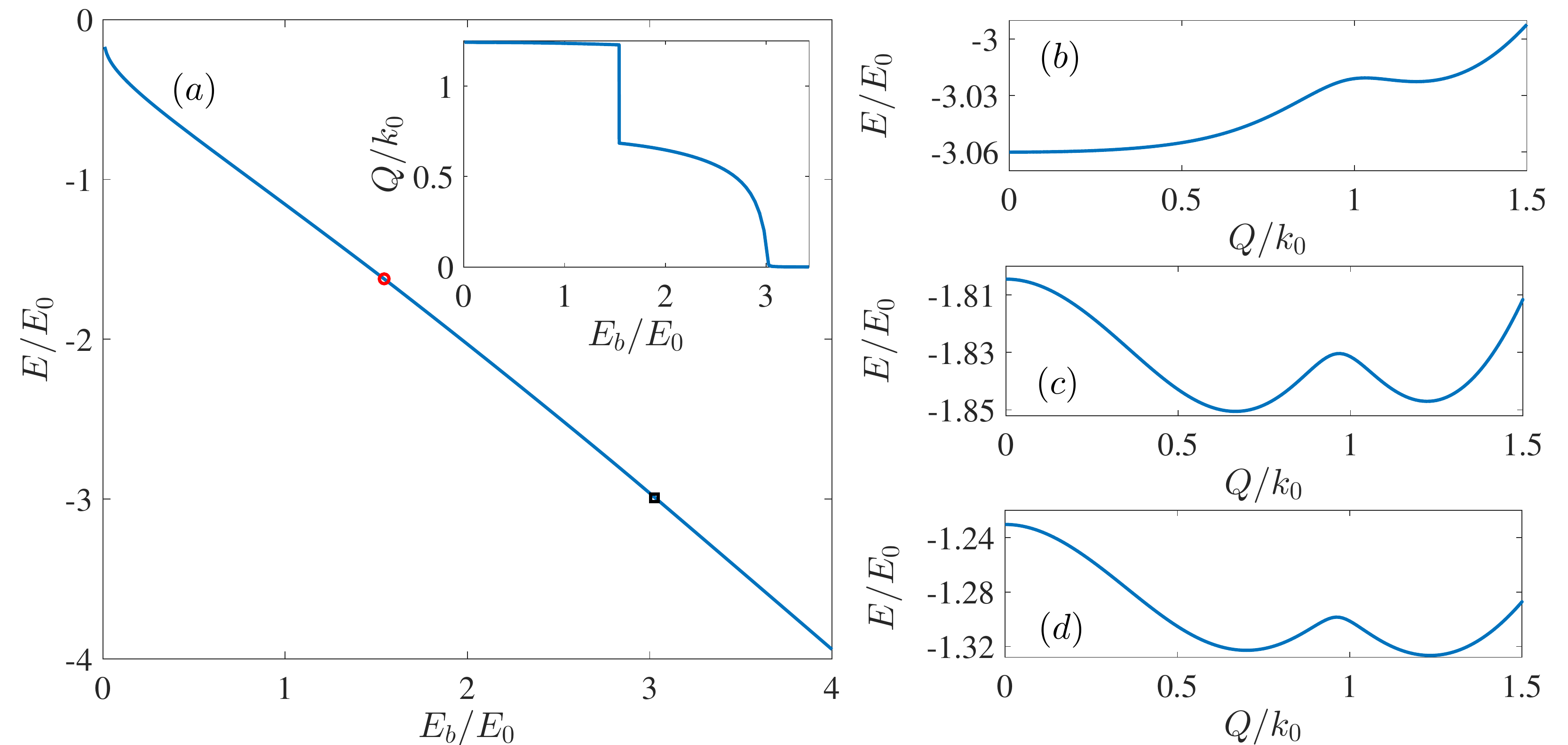}
\caption{(Color online) (a) Energy and CoM momentum (inset) of the dressed dimer with $E_F=E_{th}+ 0.05 E_0$, and the mass ratio $\eta=1$. (b)(c)(d) Dressed dimer energy as a function of CoM momentum ($Q$) for (b) $E_b/E_0=3$, (c) $E_b/E_0=1.8$, and (d) $E_b/E_0=1.2$.
}
\label{fig:suppdimer}
\end{figure*}

Another possible ground state of the system is the polaron state, which can be written as
\begin{align}
\left|P_{\cp Q}\right\rangle=\Big(\phi(\cp Q) b^{\dag}_{\cp Q} + \sideset{}{'}\sum_{\cp k\lambda\cp q\nu} \phi^{\lambda\nu}_{\cp k\cp q}(\cp Q)b^{\dag}_{\cp Q+\cp q-\cp k}a^{\dag}_{\cp k\lambda}a_{\cp q\nu}\Big) \left|FS\right\rangle_N,
\end{align}
Minimizing the energy functional, we have the equation for the polaron energy $E_P$:
\begin{equation}
E_P-\epsilon^b_{\cp Q}=\sideset{}{'}\sum_{\cp q\nu} \frac{1}{\frac{2}{U}-\displaystyle\sum_{\cp k\lambda}{}^{'}\frac{1}{E_P-\epsilon^b_{\cp Q+\cp q -\cp k}-\xi_{\cp k\lambda}+\xi_{\cp q\nu}}}.\label{eqn:polaE}
\end{equation}

\subsection{\textbf{Radio frequency spectra}}\label{supp:rf}

The r.f. spectrum of the system can be calculated following the Fermi's Golden Rule
\begin{align}
\Gamma(\omega)=\frac{2\pi}{\hbar}\sum_f |\langle f|H_{\rm rf}|i\rangle|^2\delta(\hbar\omega+E_i-E_f),
\end{align}
where the Hamiltonian for the r.f. coupling $H_{\rm rf}=V_0\sum_{\cp k} (c^{\dag}_{\cp k}b_{\cp k}+H.c.)$, $c^{\dag}_{\cp k}$ creates an atom in the bystander state with momentum $\hbar\cp k$, $i,j$ is the label for the initial and the final states, $\omega$ is the frequency of the r.f. field,
and $V_0$ is the r.f. coupling strength. The total spectrum should satisfy the condition $\int \Gamma(\omega) d\omega=1$.

For the dressed trimer
\begin{eqnarray}
\Gamma(\omega) &=& \frac{2\pi V_0^2}{\hbar}\sideset{}{'}\sum_{\cp k\lambda\cp k'\beta} |\phi^{\lambda\beta}_{\cp k\cp k'}|^2\delta(\hbar\omega+A^{\lambda\beta}_{\cp k\cp k'})+\frac{2\pi V_0^2}{\hbar}\sideset{}{'}\sum_{\substack{\cp k\lambda\cp k'\beta\\\cp k''\gamma\cp q\nu}} |\phi^{\lambda\beta\gamma\nu}_{\cp k\cp k'\cp k''\cp q}|^2\delta(\hbar\omega+A^{\lambda\beta\gamma\nu}_{\cp k\cp k'\cp k''\cp q}),
\end{eqnarray}
where the wave functions
\begin{align}
\phi^{\lambda\beta}_{\cp k\cp k'}&\propto\frac{F^{\lambda}_{\cp k}-F^{\beta}_{\cp k'}+\sideset{}{'}\sum_{\cp q\nu}G^{\lambda\beta\nu}_{\cp k\cp k'\cp q}}{A^{\lambda\beta}_{\cp k\cp k'}},\\
\phi^{\lambda\beta\gamma\nu}_{\cp k\cp k'\cp k''\cp q} &\propto \frac{G^{\beta\gamma\nu}_{\cp k'\cp k''\cp q}-G^{\lambda\gamma\nu}_{\cp k\cp k''\cp q}+G^{\lambda\beta\nu}_{\cp k\cp k'\cp q}}{3A^{\lambda\beta\gamma\nu}_{\cp k\cp k'\cp k''\cp q}}.
\end{align}

For the dressed dimer
\begin{eqnarray}
\Gamma(\omega) &=& \frac{2\pi V_0^2}{\hbar}\sideset{}{'}\sum_{\cp k\lambda} |\phi^{\lambda}_{\cp k}(\cp Q)|^2\delta(\omega+A^{\lambda}_{\cp k})+\frac{2\pi V_0^2}{\hbar}\sideset{}{'}\sum_{\cp k\lambda \cp k'\beta \cp q\nu} |\phi^{\lambda\beta\nu}_{\cp k\cp k'\cp q}(\cp Q)|^2\delta(\omega+A^{\lambda\beta\nu}_{\cp k\cp k'\cp q}),\label{eqn:rfmole}
\end{eqnarray}
where
\begin{align}
\phi^{\lambda}_{\cp k}(\cp Q) &\propto \frac{C+\sideset{}{'}\sum_{\cp q\nu}G^{\lambda\nu}_{\cp k\cp q}}{A^{\lambda}_{\cp k}},\\
\phi^{\lambda\beta\nu}_{\cp k\cp k'\cp q} (\cp Q)&\propto i \frac{G^{\lambda\nu}_{\cp k\cp q}-G^{\beta\nu}_{\cp k'\cp q}}{2A^{\lambda\beta\nu}_{\cp k\cp k'\cp q}}.
\end{align}
The spectrum of the dressed dimer is dominated by the first term of Eq.~\ref{eqn:rfmole}. From the delta function therein, it is straightforward to see that, due to the existence of a ring topology in the Fermi surface, the spectrum should have sharp edges at $\hbar\omega=-A^{-}_{\cp k_{i}}$ ($i=1,2$), where $k_1$ and $k_2$ correspond to the Fermi momenta of the inner and the outer ring of the Fermi surface, respectively. Physically, when the energy of the r.f. field $\hbar\omega$ lies within the two sharp edges, a particular branch of the out going final states is Pauli blocked by the Fermi surface.

For the polaron with zero CoM momentum
\begin{eqnarray}
\Gamma(\omega) &=& \frac{2\pi V_0^2}{\hbar}|\phi_0|^2\delta(\omega+E_P)+\frac{2\pi V_0^2}{\hbar}\sideset{}{'}\sum_{\cp k\lambda\cp q\nu} |\phi^{\lambda\nu}_{\cp k\cp q}(0)|^2\delta(\omega+A^{\lambda\nu}_{\cp k\cp q}),\label{eqn:rfpolaron}
\end{eqnarray}
where
\begin{align}
\phi^{\lambda\nu}_{\cp k\cp q}(0)&=\phi_0\frac{f(E_P,\cp q,\nu)}{A^{\lambda\nu}_{\cp k\cp q}},\\
A^{\lambda\nu}_{\cp k\cp q}&=E_P-\xi_{\cp k\lambda}+\xi_{\cp q\nu}-\epsilon^b_{\cp q-\cp k},\\
f^{-1}(E_P,\cp q,\nu)&=\frac{2}{U}-\sideset{}{'}\sum_{\cp k\lambda}\frac{1}{A^{\lambda\nu}_{\cp k\cp q}}.
\end{align}

In Eq.~\ref{eqn:rfpolaron}, the first term represents the impurity residue contribution to the polaron spectrum. While the term carries most weight of the spectrum, the delta-function form makes it difficult to be seen on the spectrum. Experimentally, finite-temperature effects and possible decay channels should broaden the delta-function and give rise to a significant peak at the polaron threshold. To reflect this, we assume a momentum-dependent Gaussian broadening in the impurity residue contribution, such that the polaron spectrum is given by
\begin{eqnarray}
\Gamma(\omega) &=& \frac{2\pi V_0^2}{\hbar}\sum_{\cp k}|\phi_0|^2\delta(\omega+E_P-\frac{\hbar^2k^2}{2m_b})\frac{1}{\pi W^2}\exp(-\frac{k^2}{W^2})+\frac{2\pi V_0^2}{\hbar}\sideset{}{'}\sum_{\cp k\lambda\cp q\nu} |\phi^{\lambda\nu}_{\cp k\cp q}(0)|^2\delta(\omega+A^{\lambda\nu}_{\cp k\cp q}),
\end{eqnarray}
where $W$ is the width of the broadening. To be specific, in the expression above, we have assumed a thermal broadening, such that $W^2=2m_bk_B T/\hbar^2$, with $T=0.05T_F$. Here we take $k_B T_F=E_F$, $k_B$ is the Boltzmann constant, and $E_F$ is the Fermi energy of a system with the same fermion density and in the absence of SOC.

For the momentum-resolved r.f. spectrum, without loss of generality, we calculate the spectrum at a fixed r.f. field frequency and a given momentum $k_x$ of the impurity atom in the bystander state.  This can be done by fixing the momentum of the impurity atom in the $x$-direction in the integration. Correspondingly, for the dressed trimer
\begin{align}
\Gamma(k_x,\omega) = \frac{2\pi V_0^2}{\hbar}\sideset{}{'}\sum_{k_y\lambda\cp k'\beta} |\phi^{\lambda\beta}_{ -\cp k-\cp k',\cp k'}|^2\delta(\hbar\omega+A^{\lambda\beta}_{-\cp k-\cp k',\cp k'})+\frac{2\pi V_0^2}{\hbar}\sideset{}{'}\sum_{\substack{k_y\lambda\cp k'\beta\\\cp k''\gamma\cp q\nu}} |\phi^{\lambda\beta\gamma\nu}_{\cp q-\cp k-\cp k'-\cp k'',\cp k'\cp k''\cp q}|^2\delta(\hbar\omega+A^{\lambda\beta\gamma\nu}_{\cp q-\cp k-\cp k'-\cp k'',\cp k'\cp k''\cp q});
\end{align}
for the dressed dimer
\begin{eqnarray}
\Gamma(k_x,\omega) &=& \frac{2\pi V_0^2}{\hbar}\sideset{}{'}\sum_{k_y\lambda} |\phi^{\lambda}_{\cp Q-\cp k}(\cp Q)|^2\delta(\omega+A^{\lambda}_{\cp Q-\cp k})+\frac{2\pi V_0^2}{\hbar}\sideset{}{'}\sum_{k_y\lambda \cp k'\beta \cp q\nu} |\phi^{\lambda\beta\nu}_{\cp Q-\cp k-\cp k'+\cp q,\cp k'\cp q}(\cp Q)|^2\delta(\omega+A^{\lambda\beta\nu}_{\cp Q-\cp k-\cp k'+\cp q,\cp k'\cp q});
\end{eqnarray}
and for the polaron with zero CoM momentum
\begin{eqnarray}
\Gamma(k_x,\omega) &=& \frac{2\pi V_0^2}{\hbar}|\phi_0|^2\delta(\omega+E_P)+\frac{2\pi V_0^2}{\hbar}\sideset{}{'}\sum_{k_y\lambda\cp q\nu} |\phi^{\lambda\nu}_{\cp q-\cp k,\cp q}(0)|^2\delta(\omega+A^{\lambda\nu}_{\cp q-\cp k,\cp q}).
\end{eqnarray}
Note that the expression for the broadened momentum-resolved polaron spectrum can be derived similar as above.

\subsection{\textbf{Applicability of the ansatz wave function}}\label{supp:validity}

We have seen that particle-hole excitations play a crucial role in stabilizing universal trimers in a spin-orbit coupled Fermi sea. In the ansatz wave functions of the dressed trimer, the dressed dimer, and the polaron, we have only considered a single pair of particle-hole excitations. For impurity problems in a Fermi sea without SOC, it has been shown previously that energies evaluated at the level of a single pair of particle-hole excitations are already quite accurate~\cite{combescot}. The fast convergence in energy with respect to the number of particle-hole excitations is due to the destructive interference between contributions within the sector of multiple particle-hole excitations. Following this argument, and using the polaron state as an example, we will show the energy of polarons with successively increasing contributions from higher-order particle-hole excitations out of a spin-orbit coupled Fermi sea also converge very fast. Therefore, we expect that ansatz wave functions with a single pair of particle-hole excitations should be sufficient to characterize dressed-state energy in a spin-orbit coupled Fermi sea.

\begin{figure*}[tbp]
\includegraphics[width=10cm]{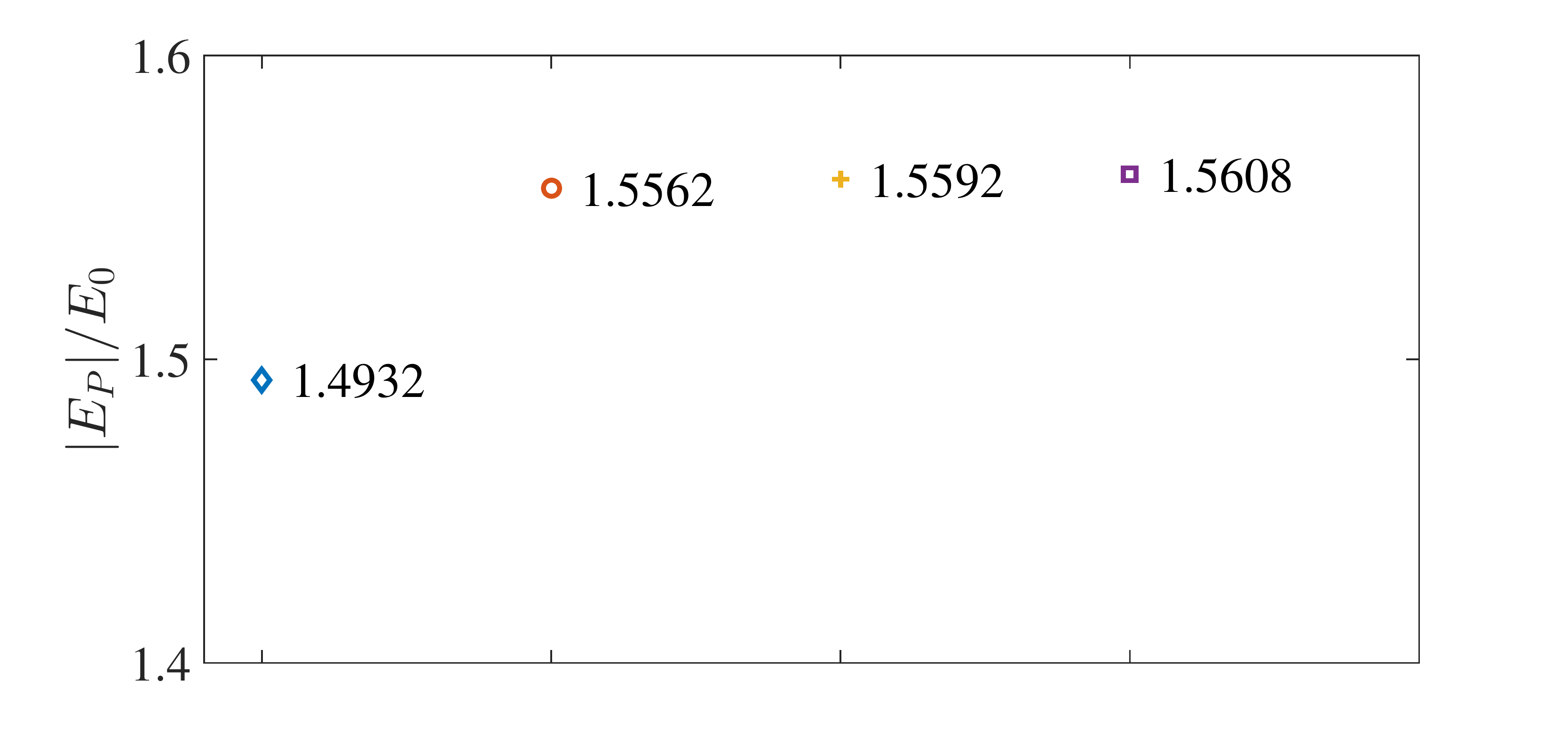}
\caption{(Color online) Comparison of the polaron energies with a single pair of particle-hole excitations (diamond), with $q=q'=k_0$ in $A^{\lambda\beta\nu\omega}_{\cp k\cp k'\cp q\cp q'}$ (circle), with $q'=k_0$ in $A^{\lambda\beta\nu\omega}_{\cp k\cp k'\cp q\cp q'}$ (cross), and with $\cp q$ and $\cp q'$ in $A^{\lambda\beta\nu\omega}_{\cp k\cp k'\cp q\cp q'}$ taking arbitrary value below the Fermi surface (square). Here, $E_b=E_0$, $E_F=E_{th}+0.2E_0$, and $\eta=1$.}
\label{fig:supp2}
\end{figure*}

To see this, we use polaron state as an example. In general, the ansatz wave function can be written as
\begin{eqnarray}
|P\rangle &=& (\varphi_0 b^{\dag}_0+\sideset{}{'}\sum_{\cp k\lambda \cp q\nu} \varphi^{\lambda\nu}_{\cp k\cp q}b^{\dagger}_{\cp q-\cp k}a^{\dagger}_{\cp k\lambda} a_{\cp q\nu}+\cdots+\sideset{}{'}\sum_{{\{\cp k_i\lambda_i\}}{\{\cp q_j\nu_j\}}} \varphi^{\lambda_i\nu_j}_{{\cp k_i}{\cp q_j}}b^{\dagger}_{\cp p}\prod^n_{i=1}a^{\dagger}_{\cp k_i\lambda_i}\prod^n_{j=1}a_{\cp q_j\nu_j}+\cdots)|FS\rangle,
\end{eqnarray}
where $\cp p = \sum^n_{j=1}\cp q_j-\sum^n_{i=1}\cp k_i$. This ansatz was first proposed for the three-dimension polaron problem in Ref.~\cite{combescot}.

Keeping terms up to two particle-hole pairs, the Schr\"odinger's equation $H|P\rangle=E_P|P\rangle$ gives rise to
\begin{eqnarray}
U^{-1}E_P\varphi_0 &=& \sideset{}{'}\sum_{\cp k\lambda\cp q\nu} \varphi^{\lambda\nu}_{\cp k\cp q},
\end{eqnarray}
\begin{eqnarray}
U^{-1}A^{\lambda\nu}_{\cp k\cp q}\varphi^{\lambda\nu}_{\cp k\cp q} &=& \varphi_0+\sideset{}{'}\sum_{\cp k\lambda}\varphi^{\lambda\nu}_{\cp k\cp q}-\sideset{}{'}\sum_{\cp q\nu}\varphi^{\lambda\nu}_{\cp k\cp q}-4\sideset{}{'}\sum_{\cp k'\beta\cp q'\omega}\varphi^{\lambda\beta\nu\omega}_{\cp k\cp k'\cp q\cp q'},
\end{eqnarray}
\begin{align}
4U^{-1}A^{\lambda\beta\nu\omega}_{\cp k\cp k'\cp q\cp q'}\varphi^{\lambda\beta\nu\omega}_{\cp k\cp k'\cp q\cp q'}~=~&-\varphi^{\lambda\nu}_{\cp k\cp q}-\varphi^{\beta\omega}_{\cp k'\cp q'}+\varphi^{\lambda\omega}_{\cp k\cp q'}+\varphi^{\beta\nu}_{\cp k'\cp q}\nonumber\\
&+4\sideset{}{'}\sum_{\cp k''\gamma}(\varphi^{\gamma\beta\nu\omega}_{\cp k''\cp k'\cp q\cp q'}+\varphi^{\lambda\gamma\nu\omega}_{\cp k\cp k''\cp q\cp q'})-4\sideset{}{'}\sum_{\cp q''
\tau}(\varphi^{\lambda\beta\tau\omega}_{\cp k\cp k'\cp q''\cp q'}+\varphi^{\lambda\beta\nu\tau}_{\cp k\cp k'\cp q\cp q''})+36\sideset{}{'}\sum_{\cp k''\gamma\cp q''\tau}\varphi^{\gamma\lambda\beta\tau\nu\omega}_{\cp k''\cp k\cp k'\cp q''\cp q\cp q'},
\end{align}
where $A^{\lambda_i\nu_j}_{\cp k_i\cp q_j}=E_P-\epsilon^b_{\cp p}-(\sum^n_{i=1}\xi_{\cp k_i\lambda_i}-\sum^n_{j=1}\xi_{\cp q_j
\nu_j})$. Importantly, these equations are formally the same as those of the polaron state in Ref.~\cite{combescot}. The main difference lies in the SOC-modified helicity branches, and hence the topology of the Fermi sea. The formal similarity ensures that a similar hierarchy of the particle-hole excitations as discussed in Ref.~\cite{combescot} also exists in our system. For example, when the dependence of $\cp q$ and $\cp q'$ in $A^{\lambda\beta\nu\omega}_{\cp k\cp k'\cp q\cp q'}$ is neglected, the equations are decoupled from contributions from higher-order particle-hole excitations. Thus, similar to the case in Ref.~\cite{combescot}, for the case of a spin-orbit coupled Fermi gas as well, an expansion in the hole wave vector $\cp q$ represents the successive inclusion of contributions from the subspace of multiple particle-hole excitations. Now, the crucial question is whether the energy of the polaron also converges quickly under SOC, when contributions from higher-order particle-hole excitations are successively included.

Defining $F^{\nu}_{\cp q}=\sideset{}{'}\sum_{\cp k\lambda}\varphi^{\lambda\nu}_{\cp k\cp q}$, $G^{\lambda\nu\omega}_{\cp k\cp q\cp q'}=4\sideset{}{'}\sum_{\cp k'\beta}\varphi^{\lambda\beta\nu\omega}_{\cp k\cp k'\cp q\cp q'}$, and neglecting some vanishingly small terms, we obtain the coupled integral equations:
\begin{eqnarray}
(\frac{2}{U}-\sideset{}{'}\sum_{\cp k\lambda}\frac{1}{A^{\lambda\nu}_{\cp k\cp q}})F^{\nu}_{\cp q} &=& \frac{\sideset{}{'}\sum_{\cp q\nu}F^{\nu}_{\cp q}}{E_P}-\sideset{}{'}\sum_{\cp k\lambda\cp q'\omega}\frac{G^{\lambda\nu\omega}_{\cp k\cp q\cp q'}}{A^{\lambda\nu}_{\cp k\cp q}},
\end{eqnarray}
\begin{eqnarray}
(\frac{2}{U}-\sideset{}{'}\sum_{\cp k'\beta}\frac{1}{A^{\lambda\beta\nu\omega}_{\cp k\cp k'\cp q\cp q'}})G^{\lambda\nu\omega}_{\cp k\cp q\cp q'} &=& -\frac{F^{\nu}_{\cp q}-\sideset{}{'}\sum_{\cp q'\omega}G^{\lambda\nu\omega}_{\cp k\cp q\cp q'}}{A^{\lambda\nu}_{\cp k\cp q}}+\frac{F^{\omega}_{\cp q'}+\sideset{}{'}\sum_{\cp q\nu}G^{\lambda\nu\omega}_{\cp k\cp q\cp q'}}{A^{\lambda\omega}_{\cp k\cp q'}}-\sideset{}{'}\sum_{\cp k'\beta}\frac{G^{\beta\nu\omega}_{\cp k'\cp q\cp q'}}{A^{\lambda\beta\nu\omega}_{\cp k\cp k'\cp q\cp q'}}.
\end{eqnarray}

Following the arguments in Ref.~\cite{combescot}, we caculate the polaron energies with a single pair of particle-hole excitations, and with two pairs of excitations. For polarons with two pairs of particle-hole excitations, we consider the following three cases: (1) $q=q'=k_0$ in $A^{\lambda\beta\nu\omega}_{\cp k\cp k'\cp q\cp q'}$; (2) $q'=k_0$ in $A^{\lambda\beta\nu\omega}_{\cp k\cp k'\cp q\cp q'}$; and (3) $\cp q$ and $\cp q'$ in $A^{\lambda\beta\nu\omega}_{\cp k\cp k'\cp q\cp q'}$ can take arbitrary values below the Fermi sea. Note that throughout our work, the Fermi surface lies in the lower helicity branch, which implies $\{\cp q,\nu\}=\{\cp q,-\}$ always. As discussed in Ref.~\cite{combescot}, these three cases represent a successive inclusion of corrections with two pairs of particle-hole excitations. As illustrated in Fig.~\ref{fig:supp2}, the polaron energy quickly converges as the approximation improves. This confirms that the Chevy-type ansatz wave functions should be applicable in our system. Due to the formal similarities of the underlying equations, we further expect that the Chevy-type ansatz should be generally applicable for impurity problems in a spin-orbit coupled Fermi sea.

\end{widetext}

\end{document}